\documentclass[multphys,vecphys]{promult}

\usepackage{makeidx}     % allows index generation
\usepackage{graphicx}    % standard LaTeX graphics tool
\usepackage{multicol}    % used for the two-column index
\usepackage{longtable}

\newcommand{\swift} {{\it Swift}}
\newcommand{\simlt}{\,\hbox{\lower0.6ex\hbox{$\sim$}\llap{\raise0.2ex\hbox{$<$}}}\,}
\newcommand{\simgt}{\,\hbox{\lower0.6ex\hbox{$\sim$}\llap{\raise0.2ex\hbox{$>$}}}\,}

\makeindex   

\begin{document}

\title*{Four Years of Realtime GRB Followup by BOOTES-1B (2005-2008)}
% \titlerunning{Short Title} % ... for short title
\author{Martin Jel\'{\i}nek\inst{1}
\and Alberto J. Castro-Tirado\inst{1}
\and Antonio de Ugarte Postigo\inst{2}
\and Petr Kub\'anek \inst{3,1}
\and Sergei Guziy\inst{1}
\and Javier Gorosabel\inst{1}
\and Ronan Cunniffe\inst{1}
\and Stanislav V\'{\i}tek \inst{4}
\and Ren\'e Hudec\inst{4,5}
\and Victor Reglero\inst{3}
\and Lola Sabau-Graziati\inst{6}
}
\institute{Instituto de Astrof\'{\i}sica de Andaluc\'{\i}a CSIC, Granada, Spain
\and Instituto Nazionale di Astrofisica, Milano, Italy
\and Image Processing Laboratory, Universitat de Valencia, Spain
\and Fakulta Elektrotechnick\'a, \v CVUT v Praze, Czech Republic
\and Astronomick\'y \'ustav Akademie v\v{e}d (AS\'U AV \v CR), Ond\v{r}ejov, Czech Republic
\and Instituto Nacional de T\'ecnica Aeroespacial, Torrej\'on de Ardoz, Madrid, Spain
}
\authorrunning{Jel\'{\i}nek, Castro-Tirado, de Ugarte Postigo et al.}

\maketitle

Four years of BOOTES-1B GRB follow-up history are
summarised for the first time in the form of a table.
The successfully followed events are described case by
case.

Further, the data are used to show the GRB trigger rate
in Spain on a per-year basis, resulting in an estimate
of 18 triggers and about 51\,h of telescope time per
year for real time triggers. These numbers grow to
about 22 triggers and 77\,h per year if we include also
the GRBs observable within 2 hours after the trigger.

\section{Introduction}
\label{sec:1}

BOOTES-1B (see also Castro-Tirado et al.
1999\,\cite{bootes1999}, Castro-Tirado et al.
2004\,\cite{bootes2004}) is an independent robotic
observatory with a 30\,cm aperture telescope dedicated
primarily to follow-up of gamma-ray burst (GRBs). Since
2003 it has used RTS2\,\cite{kubanek06} as an observing
system. It is located at the atmospheric sounding
station (Estaci\'on de Sondeos Atmosf\'ericos --- ESAt,
INTA) of El Arenosillo in Andaluc\'{\i}a, Spain (at
lat: 37$^\circ$06$'$16$''$N, long: 06$^\circ$43$'$
58$''$W). A nearby, older dome (BOOTES-1A) is used for
complementary wider angle instruments. 

% A more detailed description of BOOTES-1B has been provided
% elsewhere (cit). 
We present results of our GRB follow-up programme. In a large
table, we show a 4 year long follow-up log of BOOTES-1B GRBs ---
all triggers available in real time which were, or should have
been received and processed by the system. This selected sample
of GRBs is then used to provide a basic idea of how much time is
needed at the telescope to observe GRB optical afterglows.

% A more terse
%description is given to any GRB alert received and processed by
%BOOTES - interesting cases are described in text. The
%successful GRB follow-ups are summarized in a separate table.
%Notable events of both tables are then described in text. 

\section{Robotic Telescope Configuration}
\label{sec:2}

\begin{figure}[t!]
\begin{center}
\includegraphics[width=0.707\hsize]{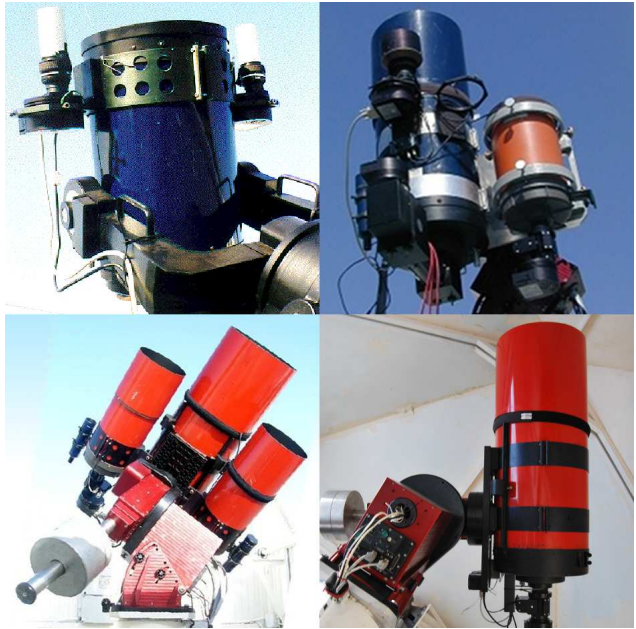}
\end{center}
\caption{Four historic BOOTES-1B configurations.}
\end{figure}

The telescope is built mostly from commercially available
components --- a Paramount ME from Software Bisque and a D=30\,cm
Schmidt-Cassegrain optical tube assembly from Meade. Over time,
four distinct system con\-fi\-gu\-ra\-ti\-ons were used,
including also two 8\,inch S-C telescopes. 

% Zminit myslenku pozorovani ze dvou stanvist.
% CGRO launched by STS- on April 5, 1991
% SAX launched 30 April 1996
% HETE-1 failed launch was on November 4, 1996
% CGRO entered atmosphere June 4, 2000
% HETE-2 was successfully launched on October 9, 2000. 
% BeppoSAX shut down on Apr. 30, 2002
% {it INTEGRAL} launched on 17 October 2002
% BepoSAX deorbited April 29, 2003,
% \swift launched on 20 November 2004
% GLAST launched June 11, 2008

\subsection{Original Meade --- stereoscopic system} 
The original BOOTES project idea of a new generation of robotic
telescopes was very simple, BOOTES-1B would --- simultaneous with
an identical setup at BOOTES-2 for parallax ability --- look for
optical transients in an extended area of the sky with wide field
cameras. Both systems would use a commercial 12-inch Meade LX-200
"robot". The wide-field cameras were considered a primary
instrument, while the ability to follow-up with a large telescope
was an option. Between 1998 and 2002 the wide-field system
provided simultaneous limits for several {\it CGRO}/BATSE and
{\it BeppoSAX} GRBs, most notably the candidate afterglow for the
short GRB\,000313\,\cite{ajct00}. The 30\,cm telescope was
successfully used to follow-up GRB\,030329. 

Although we made the original system able to observe,
it kept having problems. It required an operator
presence several times per week and, despite a notable
effort, the fork mount's electronics had to be
exchanged several times. Because of that, we decided to
purchase another mount. 

% The GRB follow-up was intended manly as an independent optical
% transient detecting instrument --- together with an identical
% setup of BOOTES-2 for parallax ability, an extended area of the
% sky would be observed by wide-field cameras to look for optical
% transients. 

% This mount turned out to be completely unsuitable for our
% needs. 

\subsection{The Prototype} 

The new incarnation of BOOTES-1B was in preparation since
mid-2002, and the first prototype was put together in November
2002 in order to follow-up {\it INTEGRAL} bursts. The most
important change was the mount to be used --- the Paramount ME
from Software Bisque. The system was still carrying a wide field
camera, but a shift had been made in priorities --- the wide field camera
performed monitoring of satellite field of view and the telescope pointed
when a trigger was received. The early stage was, however,
plagued with technical and organizational problems which
eventually delayed the first real-time real-GRB follow-up until
early 2005. 

The prototype carried three instruments on a large aluminium base
plate: The 30\,cm telescope with a field
spectrograph\,\cite{spektrograf} plus an SBIG-ST8 camera, the
20\,cm Meade (originally BOOTES-1A) telescope with an SBIG-ST9
camera observing in a fixed V-band filter, and an unfiltered
wide-field 18\,mm/1:2.8 with an SBIG ST-8 CCD (43$^\circ
\times$29$^\circ$). 

BOOTES-1B was operating with this setup for about a year --- on
June 2004 it was dismounted and sent to the Workshop in
Ond\v{r}ejov for a definitive solution.

% There is one, untold discovery of \swift satellite - In HETE
% times, it was generally believed that the OT would rise until
% few tens of seconds after the GRB and then decay. If such a
% behavior is back-extrapolated from the known optical
% transients, the magnitudes expected are of the order of 5-10. 
%
% Followup of \swift GRBs has shown, that the early afterglows
% are rarely brighter than mag. 12.

\begin{figure}[t!]
\begin{center}
\includegraphics[width=\hsize]{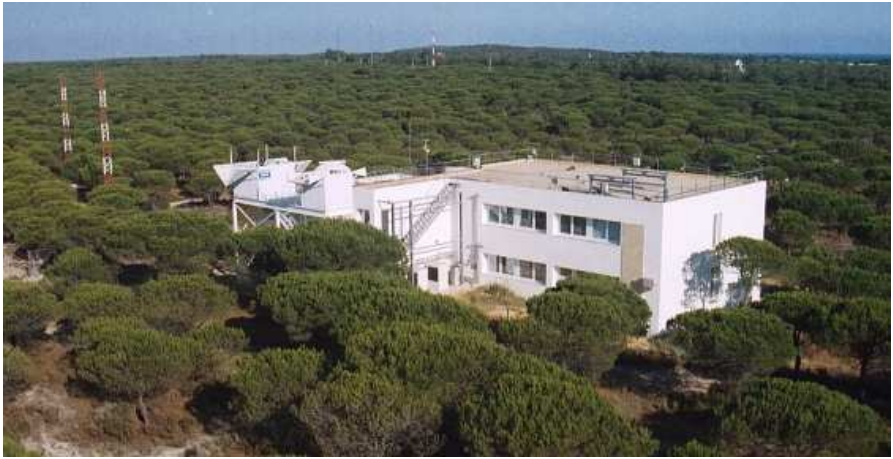}
\end{center}
\caption{The building of ESAt with domes of BOOTES-1A
and BOOTES-1B (2002).}
\end{figure}

\subsection{Triple Telescope} 

The prototype was very heavy and from the beginning had some
problems. In September 2004 BOOTES-1B finally received an upgrade
--- together with the 30\,cm telescope, there were also two
20\,cm telescopes\footnote{One of them lent personally by AJCT} for direct imaging in different filters. The
system was completely redesigned with many mechanical
improvements and was built to be as light as possible to allow
the mount working at its maximum slewing speed. In belief that
the rapid dissemination and fast followup after the launch of
\swift\/ would lead into relatively frequent detections of bright
optical counterparts, the 30\,cm telescope was equipped with a
field spectrograph and two 20\,cm telescopes with fixed V\&I-band
filters. The limiting magnitude of all three instruments was V
$\sim$ 16 for a 60\,s exposure. The wide-field cameras were moved
from BOOTES-1B to BOOTES-1A.

Later, during the telescope operation it became clear that the
GRB optical transients were not as bright as had been expected
and so the spectrograph on the 30\,cm telescope was replaced with
a direct imaging CCD with R-band filter --- improving the
limiting magnitude but losing the spectroscopic ability. 

On April 23, 2006, The ESAt building was struck by lightning
during a storm, destroying a major part of BOOTES-1B electronics.
It took more than a year to get BOOTES-1B definitively back
online.

\subsection{Single 30\,cm telescope}
 
During the lengthy reconstruction of BOOTES-1B, the followup
strategy was reconsidered: in the interest of detecting more
optical transients the filter(s) were abandoned
($\sim$2.5$\times$ or 1 magnitude gain in sensitivity). The
limiting magnitude of an unfiltered 120\,s exposure would be
about 18.0 --- effectively doubling the likelihood of getting an
optical transient in comparison with the R-band imaging (cf.
Fig.\,\ref{bright}).  Both 20\,cm telescopes were dropped because
of lack of suitable CCD cameras available for them. Since then,
BOOTES-1B has only a single 30\,cm telescope. 

Any observations obtained after June 15, 2007 have been
obtained without filter (W for white). We calibrate
them against R-band, which, in in the case of no color
evolution of the optical counterpart, is expected to
result in a small constant offset in magnitude.

\section{Real Time GRB followup}

BOOTES-1B could have received during the past 4 years
(since January 2005 until December 2008) 86 GRB
triggers via GCN, which could have been followed in
real time or would become observable within the
following two hours.  Table~\ref{bigtable} summarizes
these triggers, noting, among BOOTES-1B status of the
followup, also the brightness of the GRB optical
counterpart, if it is known. The magnitude estimation
search was done with a heavy use of
GRBlog\,\cite{grblog}. We use these data to construct a
"limiting magnitude vs. likelihood of detection" graph
(Fig.\,\ref{bright}).

\subsection{GRB triggers followed by BOOTES-1B}

\paragraph{GRB\,050215B}

GRB was discovered by \swift/BAT at 02:33:43.2\,UT.
BOOTES received the notice, 
% but failed to compute well the local horizon and
but because of a software error waited with the slew
until $\sim$22\,minutes after the GRB.  We coadded
600\,s exposures taken by both 20\,cm telescopes to
obtain limits of V$>$16.5 and I$>$15.0\,\cite{gcn3023}.

\paragraph{GRB\,050505}

First image of this \swift-discovered
GRB\,\cite{gcn3360} was obtained at 23:32:30\,UT, i.e.
609\,s after the trigger and 70\,s after receiving the
coordinates. No optical afterglow was
detected\,\cite{gcn3376}.

\paragraph{GRB\,050509A}

At the time of this trigger, the dome was still
operating independently on the rest of the system - we
obtained the first image 23\,s after the GRB trigger
(6\,s after receiving the alert). The dome was closed
due to what we consider a false trigger on the rain
sensor. The first useful 10\,s exposure was obtained
63.8\,min after the burst and has a limiting magnitude
of V$>$14.9. A coadd of first 112$\times$10\,s
exposures with an exposure mean time 88.0\,min after
the GRB has a limit of V$>$18.1.

\paragraph{GRB\,050509B}

This was a \swift-detected short gamma-ray
burst\,\cite{g050509b}. Starting 62\,s after the
trigger, BOOTES-1B seems to have obtained the
world-first data set of this short duration GRB.
However, bad luck caused that the location of the GRB
on the sky coincided with the tip of a nearby antenna
and it's signalling light.  The limiting magnitude is
thus seriously degraded. The first 10\,s exposure has a
limiting magnitude V $>$ 11.5, a combination of the
first 12$\times$10\,s exposures provides V$>$12.5. 

% 050509B B-1B r 62s 48s \swift-SHORT Antenna, SLIZ (V>~12.5 pro
% 62-296s post, ~11.5 10s exposure 62s-72s post-burst)

\paragraph{GRB\,050525A}

GRB\,050525A\,\cite{gcn3466} was the first BOOTES-1B burst for
which a detection was obtained. The telescope started the first
exposure 383\,s after the GRB trigger (28\,s after receiving the
notice). An optical afterglow with V$\simeq$15.0 was detected.

% 050525 B-1B r 383s 28s LC, ~15.0 na zacatku, Jen V

\paragraph{GRB\,050528}

We observed the errorbox\,\cite{gcn3496} at 04:07:56\,UT, ie.
starting 71\,s after the burst and 28\,s after receiving the
trigger in a light twilight, setting the limit to the possible
GRB counterpart to V$>$13.8 and I$>$13.0 during the first 60\,s
after the beginning of our observation\,\cite{gcn3500}.

% 050528 B-1B r 71s 28s \swift V>13.8, I>13.0

\paragraph{GRB\,050730}

Located at 19:58:23\,UT, this GRB was very low above horizon in
real time, BOOTES-1B obtained few exposures starting 233.4\,s
after the GRB, when the system failed. The images did not provide
detection of the ~17.0\,mag optical transient discovered by both
\swift/UVOT\,\cite{gcn3717} and 1.5\,m telescope at
OSN\,\cite{shashi}.

\paragraph{GRB\,050805B}

This short burst was localized by \swift\/ at
20:41:26\,UT, BOOTES obtained first images 62.2\,s
after the trigger (7.2\,s after receiving the trigger).
No optical transient was detected. The first 10\,s
exposure has a limiting magnitude R$>$16, combination
of first five images (exposure mean time 118\,s after
the GRB) has a limit R$>$17.0.

\paragraph{GRB\,050824}

The optical afterglow of this GRB was discovered with
the 1.5\,m telescope at Sierra
Nevada\,\cite{gcn3866,jesper}. BOOTES-1B was, however,
the first telescope to observe this optical transient,
starting 636\,s after the trigger with R$\simeq17.5$.
The weather was not stable and the focus not perfect,
but BOOTES-1B worked as expected.  Eventually, several
hours of data were obtained.

\paragraph{GRB050904}

BOOTES-1B reacted to this GRB, starting 124\,s after
the trigger.  There was a hot pixel close to the GRB
location, which made us believe we might have a
detection in the {\it R}-band, which was issued in the
first BOOTES-1B circular. Later, the observation
revised as a limit (R$>$18.2) which was used to compute
the record redshift of this GRB\,\cite{nature1}. The
I-band camera of the 20\,cm telescope, unluckily,
failed. 

\paragraph{GRB\,050922C}

This bright burst was detected by \swift/BAT at
19:55:50\,UT.  BOOTES was not very lucky - the weather
on the station was bad.  Instead of a limiting
magnitude of $\sim$ 17.0 for a 30\,s exposure we got
12.9.  The afterglow was eventually detected with the
R-band camera (at the 30\,cm telescope) on few
occasions between flying clouds. The first weak
detection was obtained 228\,s after the GRB trigger
gave R$\simeq$14.5. 

\paragraph{GRB\,051109A}

This is the only GRB ever detected by BOOTES-1B
simultaneously in more than one filter. The first
images were obtained 54.8\,s after the burst in R and I
bands\,\cite{gcn4227}.

\begin{figure}[t!]
\begin{center} \includegraphics[width=0.870\hsize]{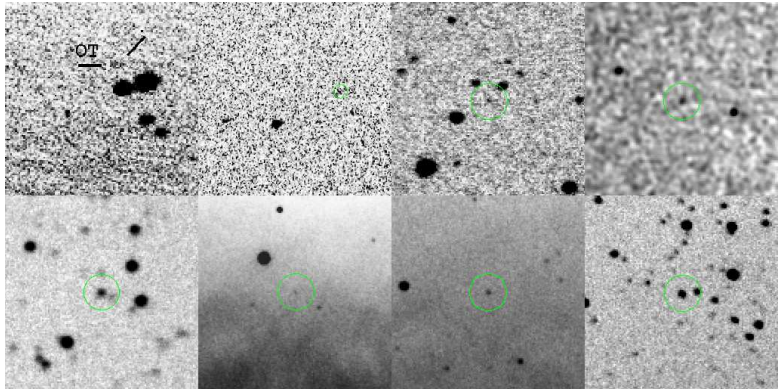}
\end{center}
\caption{GRB optical transient detections by BOOTES-1B: first
row: GRB\,050824, GRB\,050922C, GRB\,051109A, GRB\,080330. Second
row: GRB\,080413B, GRB 080430, GRB\,080602B, GRB\,080605. }
\end{figure}

\paragraph{GRB\,051211B}

Observation of this burst started 42\,s after the burst. A 30\,s
R-band exposure was obtained, but the camera failed after getting
this image. Only useless defocused I-band images were taken with
the 20\,cm telescope\,\cite{gcn4333}.

\paragraph{GRB\,051221B}

This GRB was detected by \swift/BAT at
20:06:48\,UT\,\cite{gcn4376}.  BOOTES-1B slewed to the position
and started obtaining images 27.8\,s after receiving the alert
(234.8\,s after the burst). We did not find any new source in our
images\,\cite{gcn4379}. 30 min after the trigger, a faint 21
magnitude afterglow was discovered elsewhere\,\cite{gcn4381}.

%\paragraph{GRB\,060111}
%
%The GRB was localized by Swift during BOOTES-1B
%maintenance (CCD focussing). Resulting images were
%therefore not perfectly focused and the limit was
%worsened. XXX

\paragraph{GRB\,060421}

BOOTES-1B reacted to this GRB within 61\,s after the
trigger. Images were not of a great quality, yielding a
limit of R$>$14 for the first 10\,s exposure and R$>$16 for
the combination of 30 images (exposure mean time 547\,s
after the trigger).

\paragraph{GRB\,061110B}

The GRB was detected by \swift/BAT at T$_0=$21:58:45\,UT,
but the notices were delayed by 626\,s because \swift\/ was
performing downlink.  BOOTES-1B started to slew immediately after
reception of the trigger, obtaining the first image 698\,s after
the burst (72\,s after the GCN notice). When seven 60\,s
exposures were obtained, a communication error with the mount
occurred. The communication was later restored and further 19
images were obtained starting 22:49:49\,UT (0.85\,h after burst).
Last image was obtained at 23:36:20 (1.62\,h after burst). 

Combination of the first 7 images (limiting magnitude
$\sim$17.2\,mag each) with the exposure mean time 938\,s after
the GRB trigger was found to have a magnitude limit $\sim$18.0.
Combination of 11 images obtained between 22:49:49\,UT and
23:03:45\,UT (lim $\sim$ 16.9\,mag each) yields a limit of
$\sim$18.2 with mean time T$_0+$3452\,s.

\paragraph{GRB\,071101}

% 54.8s 23.3s S295779 GCN 7032: W>17.0 (T50038)

First successful followup of a GRB after installation of the
30\,cm telescope in BOOTES-2 site at La Mayora. The GRB
trigger\,\cite{gcn7030} was at 17:53:46\,UT. BOOTES started
imaging 54.8\,s after the burst (23.3\,s after receiving the
coordinates). No afterglow was detected, an unfiltered, R-band
calibrated limit of W$>$17.0 was estimated\,\cite{gcn7032}. 

\paragraph{GRB\,071109}

{\it INTEGRAL} detected this GRB at 20:36:05\,UT\,\cite{gcn7046}.
BOOTES followed up 58.5\,s after the GRB (30.9\,s after receiving
the alert).  Because of high altitude clouds, the telescope
performance was reduced, yielding an unfiltered limit of
$\sim$13.0 in the first 10\,s exposure\,\cite{gcn7047}.

\paragraph{GRB\,080330}

This GRB\,\cite{gcn7537} happened during the first day
recomissioning of BOOTES-1B after its move from
BOOTES-2 site in La Mayora. The GCN client was not yet
operational and at the time of the GRB we were focusing
the telescope. First image was obtained 379\,s after
the GRB trigger and the optical afterglow was detected
with magnitude $\sim$16.3 on the first image. A bug in
the centering algorithm caused a loss of part
subsequent data. Further detections were obtained
starting 21\,min after the GRB when the problem was
fixed.

\paragraph{GRB\,080413A}

BOOTES-1B started obtaining images of the
GRB\,080413A\,\cite{gcn7603} starting 60.7\,s after the trigger
(46.3\,s after reception of the alert). A W$\simeq$13.3
magnitude optical afterglow was found\,\cite{mates080413A}. 

% 60.7s 46.3s S309096 Lightcurve! (W~13.5) GCN7603 (T50082)

\paragraph{GRB\,080430}

BOOTES-1B obtained the first image of this GRB\,\cite{gcn7648}
34.4\,s after the trigger. An optical transient was found with a
magnitude W $\simeq$ 15.5\,\cite{adup080430}.

% 34.4s 22.1s S310613 Lightcurve! (W~15) GCN7648 (T50090) At
% twilight sky

% Together, 321 images of 6,20 and 60\,s were obtained. Total
% exposure time were 4\,h. BOOTES-1B was able to detect the OT
% until ~30\min after the GRB.

\paragraph{GRB\,080603B}

This GRB happened at BOOTES-1 site during sunset. We obtained
first useful images starting one hour after the trigger. An W
$\simeq$ 17.4 optical transient was found. 

% 41.9s 26.2s S313087 60m after the GRB a first usable image.
% T50113 Not real time observation

\paragraph{GRB\,080605}

GRB\,080605 was observed starting 41.9\,s after the trigger. A
rapidly decaying optical afterglow with W $\simeq$ 14.8 was
found\,\cite{gcn7837}.

% 41.9s 29.3s S313299 R~15, LC, GCN7837, T50115

\paragraph{GRB\,081003B}

{\it INTEGRAL} detected this GRB at 20:36:05\,UT\,\cite{gcn8317}.
BOOTES started obtaining unfiltered images at 20:48:49\,UT (41\,s
after the GRB trigger and 17.4\,s after the GCN notice), single
images have a detection limit of W $>$ 14\,mag. The combination of
the first 32 images with an exposure mean time of 80\,s after the
GRB has a limit of W $>$ 17.6\,mag (calibrated against GSC2).
Neither shows any new sources within the GRB
errorbox\,\cite{gcn8320}.
 
\begin{table} \begin{center} 
{ \small
\begin{longtable}{lccl}
\hline
\hline
GRB & 
\quad$T_{\mathrm{obs}}-T_{\mathrm{trigger}}$ \quad& 
\quad$T_{\mathrm{obs}}-T_{\mathrm{notice}}$ \quad& 
mag \\ % note \\
%\qquad& $T_{\mathrm{start}}-T_0$ & $T_{end}-T_0$ & magnitude & note\\
\hline
050215B& 22\,min&       &        V$>$16.5, I$>$15.0\\
050505&  609\,s&        70\,s&   V$>$14.0\\ % a rough estimate
050509A& 63.8\,min&     &        V$>$18.1\\
050509B& 62\,s&         48\,s&   V$>$12.5\\
050525A& 383\,s&        28\,s&   V$\simeq$15.0\\[4pt]
050528&  71\,s&         28\,s&   V$>$13.8, I$>$13.0\\
050730&  233\,s&        172\,s&  R$>$16\\
050824&  636\,s&        55.8\,s& R$\simeq$17.5\\
% 050801   dt=47s
% 050803   dt=1.77h
% 050805A  dt=136s
050805B&  62\,s&        17\,s&	 R$>$16.0\\
% 050815   dt=4h
% 050822   dt=271s
050904&  124\,s&        43\,s&   R$>$18.2\\[4pt]
050922C& 228\,s&        62.3\,s& R$\simeq$14.5\\
051109A& 54.8\,s&       27\,s&   R$\simeq$16.2\\
051211B& 42\,s&         48.4\,s& I$>$14.0\\
051221B& 234.8\,s&      27.8\,s& V$>$13.3\\
%060111A& 296\,s&	&	 R$>$XXX\\  
060421 & 61.2\,s&       47.6\,s& R$>$16.0\\[4pt]
061110B& 698\,s&        72\,s&   R$>$18.0\\
071101&  54.8\,s&       23.3\,s& W$>$17.0\\
071109&  58.5\,s&       30.9\,s& W$>$13.0\\
080330&  379\,s&        &        W$\simeq$16.3\\
080413A& 60.7\,s&       46.3\,s& W$\simeq$13.3\\[4pt]
080430&  34.4\,s&       22.1\,s& W$\simeq$15.5\\
080603B& 60\,min&       &        W$\simeq$17.4\\
080605&  41.9\,s&       29.3\,s& W$\simeq$14.8\\
081003&  41\,s&         17.4\,s& W$>$14.8\\
\hline
\hline
\end{longtable}
\caption{Summary of GRBs successfully followed by BOOTES-1B. }
\label{followup_tbl}
}
\end{center} \end{table}

\section{Implications}

% \subsection{Weather}
% 
% General weather statistics of BOOTES-1B can be derived. In this
% GRB-selected study the weather was good only in 19 cases of 45
% which were successfully followed (42\,\%). In 3 cases (7\,\%) the
% images were taken through clouds and in 23 cases (51\,\%) the
% weather did not allow observation. 
% 
% This weather statistic is a little biased by the fact that in
% summer, the night length is about 50\,\% of the winter night. The
% rainy weather in winter is therefore having more influence to the
% overall image (traditional statistics give 75\,\% of clear
% skies). 
% 
%the summer when the statistic is always best. However, in summer
%the nights are shorter, so the number of the realtime GRBs is 
%
%With about 40\,\% of bad weather the weather at BOOTES actually
%appears much worse than it really is - this is caused by the
%fact that we have lost 3 summers because of major system
%failures. 

\subsection{Success rate}

Of the 89 triggers, 45 were processed in realtime and
observed if possible, in 44 cases the system could not
respond. This makes the overall failure rate quite high
(50\,\%). 29 triggers were, however, lost due to
long-term failures resulting from the telescope being
struck by lightning. 8 more triggers failed during
first 6 months of operation, when the system was not
yet fully stable and one was lost during
maintenance (and followed manually). 6 triggers out of
47 (13\,\%) were lost unexpectedly during the 963 nights of
telescope operation if we do not count the first
semester of 2005.

% Of the triggers which could have been observed, about one third
% failed to be executed because of various failures - a
% particularly painful is the failure of the link to the GCN,
% caused by an unreliable Internet connection. Other hardware and
% software errors were also experienced in some cases. Overall
% reliability of the system is about 66\,\%, but most of the
% problems happened during the first year of operation.
% 
\subsection{Planning} 

When specifying the GRB follow-up needs, the number of nights
(hours) spent observing GRBs has to be estimated. Under various
follow-up strategies we may derive different results.

Due to various instrumental effects (like a passage through the
South Atlantic Anomaly) related to the satellite \swift, an
offset from the overall triggering statistic which would depend
on a geographical location could be found.

In Table\,\ref{bigtable}, the fourth column has the time in hours
until the first set of the event location below 10$^\circ$ of
altitude or until the Sun rises above -15$^\circ$ of altitude.
For non-realtime events, this is the time the location spends on
the night sky, for real-time triggers, it is the time between the
trigger and the moment when the target becomes unobservable. 

For a small telescope, we assume that once the GRB is real-time
triggered, it is unlikely to detect it the following night (i.e.
after $\sim$24\,h), so we assume the following simple follow-up
strategy: Let the telescope observe the GRB once it becomes
accessible for the first time (which is immediately for real-time
triggers) and let it observe until the GRB sets or the night
ends. Do not observe any further nights. Under the given
assumptions we get the following observing needs (assuming
perfect weather):

\paragraph{Real time triggers}

There have been 72 real time triggers during the studied 4 years,
during their first nights they accumulated 202 hours.

So if we allow only real-time followable triggers to be observed,
we would need $\sim$ 18 triggers per year (once per 20 days) and
on average 2.8\,h (max.  8.0\,h) of observing time per trigger,
50.5\,h per year. Such a program would consume about 2\,\% of
the telescope time.  

%derive the need as ... hours per year.  In this set, there were
%46 triggers followable in realtime during 1079 days, i.e. $\sim$
%7.8 per year.  i.e. one trigger per 18.6 days or $\sim$9.8
%triggers per year. When we include also the triggers followable
%within 2 hours after the GRB, we get a total of 46 triggers,

\paragraph{Extended set}

In the extended set, we assume that GRBs that would become
observable within 2 hours after the event would also be followed.
We would need $\sim$22 triggers per year, each with an average
length of 3.5\,h. In total we would need 78.5\,h per year, or
about 3\,\% of the telescope time.

\begin{center} 
%\begin{longtable}[b!]
%
{ \small
\begin{longtable}{lcc}
\hline
\hline
 & \qquad Real time only \qquad& \qquad Up to 2\,h \qquad \\
\hline
triggers/year & 18& 22\\
hours/year & 50.5 & 78.5\\
hours/trigger & 2.8 & 3.5 \\
days/trigger & 20.3 & 16.6 \\
%\hline
\hline
\hline
% 090715B&	\swift-
%
\\
\caption{Results of the GRB-planning statistic}
\end{longtable}
}
%\end{longtable}
\end{center}

\subsection{Optical Afterglow Brightnesses}  

% sync this w/ image caption!  mp: P.6 - mozna by melo byt i v
% textu, ne jen v popisku obrazku, ze ten vypocet vychazi z
% Tabulky 1.

\begin{figure}
\begin{center}
\includegraphics[width=0.600\hsize]{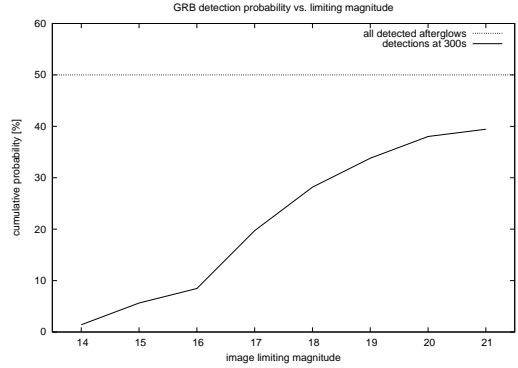}
\end{center}
\caption{The graph (based on $T_{300}$ data from
Table\,\ref{bigtable}), showing the likelihood of detection of an
optical afterglow of a GRB as a function of the magnitude the
telescope can detect (in the time interval discussed here). The
dotted line delimits 50\,\% --- the ratio of GRBs in our data for
whose there was eventually discovered an optical transient.}
\label{bright}
\end{figure}

As a representative value of GRB optical transient brightness,
important for real-time follow-up, we have chosen its magnitude
at 300\,s after the trigger.  It turns out that it is not easy to
find a uniform sample and available magnitudes and limits are a
mixture of different passbands, mainly V,R and unfiltered CCD
magnitudes. For a general idea of how bright an OT could be this
is, however, good enough.  Fig.\,\ref{bright} shows a cumulative
probability of detecting an OT five minutes after the trigger
with a telescope able to detect a given magnitude. For many GRBs
the brightness at this early time is unknown, or only a limit
from small telescopes has been established, so this curve is
actually a slight underestimation. 

For example BOOTES-1B, which could detect mag $\sim$ 18 at an
unfiltered 60\,s exposure, may detect an OT in about one third of
the GRB triggers. 

%Bude potreba znovu hledat reference k jasnostem GRBu v tabulce.
%:|

% \dag - raw estimates by $\alpha*2.5*\log(t/t_{0})$ against
% reference, $\alpha$=1.0 if unknown.

% denote which B-1B setup was used for which followup make an
% extra column with a B-1B result "V>12", "R>5" etc.
%
\section{Conclusions} \label{sec:5}

We have shown in a small historical retrospective the evolution
of the telescope of BOOTES-1B, as it developed from the
wide-field survey telescope to a dedicated GRB follow-up
telescope. 

Four years of BOOTES-1B GRB follow-up history are
summarised for the first time transparently in the form
of a table, which includes not only the observation
status of BOOTES-1B, but also the time for which the
object could have been observed, and a magnitude (or a
limit) of the GRB optical afterglow 5 minutes after the
trigger found in the literature. Every existing GRB
trigger which was, or could have been observed by
BOOTES-1B within 2 hours after the trigger is included. 
%
%The data present in the table are used to derive the
%observing conditions of the observatory and its
%reliability. The weather, in contrast to the behaviour
%expected and measured by other means (95\% clear
%nights between June and September, 70\% all year)
%seems poor, but evidently, this bias is caused by two
%summers being lost because of major system failures. 
%
Twenty four successfully followed events are described
case by case in a separate chapter. Many of these are
published for the first time. 

The data collected are also used to show the GRB
trigger rate in Spain. By simply counting the triggers
and the days during which they were collected, we
estimate 18 triggers and about 50.5\,h of telescope
time per year for real time triggers. These numbers
grow to about 22 triggers and 78.5\,h per year if we
include also the GRBs observable within 2 hours after
the trigger. We also derive the likelihood of the
optical afterglow detection five minutes after the GRB
trigger depending on the limiting magnitude of the
telescope.

%four years of GRB operational history of BOOTES-1B are
%summarized in this text.  Tables can be used to determine the
%actual need for telescope time for GRB follow-up etc.

%This paper is authoritative in the sense that the authors did
%their best effort to include every single burst followable by
%BOOTES-1B since its first light until May 2009. It is unlikely
%that further investigation would reveal better results. 

\begin{center} 
%\begin{longtable}[b!]
%
{
\small%\sf
%\scriptsize
\begin{longtable}{lcccccl}
\hline
\hline
Object & Target & $t_1$[h] & $t_{obs}$[h] & $m_{300}$[mag] & $dT$ &	observation status \\
%\hline
\hline
050128 &	-&	+0.0&	1.8&	-&				-&	No link to GCN\\
050208 &	-&	+0.0&	4.2&	-&				-&	No link to GCN\\
050215B&	5064&	+0.4&	10.0&	$\simgt16^\dag$\cite{gcn3022}&	22\,m&	V,I limits \cite{gcn3023}\\
050306&		5075&	+0.2&	1.8&	$>$16\cite{gcn3107}& 		86\,s&	w/roof closed\\  
050416B&	5109&	+0.0&	2.8&	-&				-&	{\tt grbd} failure \\ 
050421&		5112&	+0.0&	0.2&	$>$18.4 \cite{gcn3304}&		-&	hw problems\\ 
050502A&	-&	+0.0&	1.8&	16.3\cite{guidorzi05}& 		-&	{\tt grbd} failure\\ 
050505&		5123&	+0.0&	2.2&	-$^\dag$&			609\,s&	clouds, V,I limits\\ 
050509A&	5129&	+0.0&	2.2&	$>$18.2\cite{gcn3394}&		23\,s&	hw problems, later limit\\
050509B&	5130&	+0.0&	0.0&	$>$20.8\cite{gcn3414}&		62\,s&	OK, $V>12.5$, antenna hit!\\
050520&		-&	+0.0&	3.6&	$>$16.6&			-&	GCN connection lost\\ 
050525A&	5136&	+0.0&	3.6&	14.7\cite{blustin06}&		383\,s&	OK, V-band lightcurve \\ 
% 050528&	5137&	+19.8&	3.6&	$>$17.2\cite{gcn3497}&		71\,s&	OK, limits \cite{gcn3500} \\
050714A&	1037&	+0.0&	3.6&	-$^\dag$&			10\,m&	manually, later limit\\ 
050730&		50008&	+1.2&	1.6&	17.4\cite{gcn3717}&		1\,h 40\,m&	limits \\ 
% 050801   dt=47s	+2.6&   0.0&    
% 050803   dt=1.77h
% 050805A  dt=136s      +2.8&   7.2&
050805B&	50015&	+0.2&	7.2&	-& 62\,s&	limits\\
% 050815   dt=4h
% 050822   dt=271s (40m pred koncem noci, 3-4.5 stupne nad obzorem)
050824&		50032&	+0.0&	5.2&	17.5\cite{jesper}&		636\,s&	detection \cite{jesper} \\ 
050904&		50055&	+0.0&	2.8&	-$^\dag$&			124\,s&	R-band limit \cite{nature1}\\ 
050922C&	50090&	+0.0&	6.2&	15.5\cite{gcn4041}&		228\,s&	detections between clouds \\ 
051109A&	50126&	+0.0&	1.4&	16.8&				54.8\,s&detection in R,I\\ 
051111&		-&	+0.0&	4.6&	14.9\cite{gcn4250}&		-&	GCN connection lost\\ 
051211A&	50144&	+0.0&	3.2&	-$^\dag$&			-&	CCD failure\\ 
051211B&	50146&	+0.0&	4.8&	$>$14.0&			50\,s&	OK, limits \cite{gcn4333} \\ 
051221B&	50151&	+0.0&	3.8&	$>$18.2\cite{gcn4386}&		235\,s&	OK, limits \cite{gcn4379} \\ 
051227&		50155&	+1.6&	10.6&	$>$19.2\cite{gcn4411}&		59\,m&	bad weather\\ 
060111A&	50162&	+0.0&	2.0&	$>18.3$\cite{gcn4483}&		296\,s& during maintenance, limit\\ 
060121&		-&	+0.0&	7.8&	-$^\dag$&			-&	telescope OFF\\ 
060123&		50171&	+0.0&	7.6&	-&				-&	bad weather\\ 
060130&		50173&	+0.0&	1.2&	-&				-&	bad weather\\ 
060203&		-&	+0.0&	6.2&	-&				-&	bad weather+no GCN\\ 
060204C&	-&	+0.0&	9.6&	$>18.7$\cite{gcn4665}&		-&	bad weather+no GCN\\ 
060206&		-&	+0.0&	1.4&	16.5\cite{gcn4693}&		-&	GCN connection lost\\ 
060219&		50185&	+1.0&	6.0&	$>$18.6\cite{gcn4798}&		-&	bad weather\\ 
060319&		50190&	+0.0&	4.4&	$>$19$^\dag$\cite{gcn4889}&	-&	bad weather\\ 
060418&		50207&	+0.0&	1.4&	14.2\cite{gcn4976}&		-&	bad weather\\ 
060421&		50208&	+0.0&	3.8&	$>$16.8\cite{gcn4988}&		61.2\,s&limit\\ 
% \hline
060424&		-&	+0.0&	0.0&	-$^\dag$&			-&	hw failure\\ 
060502A&        -&      +0.0&	1.0&    18.7\cite{gcn5068}&             -&      hw failure\\ 
060507&         -&      +0.0&   2.0&    $>15.5$$^\dag$\cite{gcn5092}&   -&      hw failure\\ 
060512&         -&      +0.0&   4.6&    17.15\cite{gcn5130}&            -&      hw failure\\ 
060515&         -&      +0.0&   1.4&    $>$16.2\cite{gcn5134}&          -&      hw failure\\ 
060522&         -&      +0.0&   1.6&    19.65\cite{gcn5158}&            -&      hw failure\\ 
060602A&	-&	+0.0&	1.2&	$>$15$^\dag$\cite{gcn5201}&	-&	hw failure\\ 
060602B&	-&	+0.0&	3.6&	-&				-&	hw failure\\ 
060712 &	-&	+0.2&	3.0&	$>14.5$\cite{gcn5303}&		-&	hw failure\\ 
060814 &	-&	+0.0&	0.8&	$>$17.4\cite{gcn5448}&		-&	hw failure\\ 
060825 &	-&	+0.0&	1.4&	$>$18.3\cite{gcn5476}&		-&	hw failure\\ 
060901 &	-&	+1.6&	5.2&	-$^\dag$&			-&	hw failure\\ 
060904A&	-&	+0.0&	0.2&	$>$19.5\cite{gcn5506}&		-&	hw failure\\ 
060904B&	-&	+0.0&	2.2&	$\sim$17\cite{rykoff09,mj694b}&	-&	hw failure\\ 
% \hline
060929&		50212&	+0.0&	1.2&	$>$17.0\cite{gcn5658}&		-&	bad weather\\
061019&		50220&	+0.0&	1.0&	$>$14.8$^\dag$\cite{gcn5731}&	-&	bad weather\\
061110B& 	50228&	+0.0&	2.2&	$>$17.8$^\dag$\cite{gcn5808}&	11\,m38\,s&	$>18.1$ OK \\ 
% \hline
061217 &	50240&	+0.0&	2.6&	$>$19.2\,\cite{gcn5942}&	-&	mount failure\\ 
061218 &	50242&	+0.0&	2.0&	$>$18.6\,\cite{gcn5937}&	-&	mount failure\\ 
061222B&	50245&	+0.0&	0.6&	18.0\,\cite{gcn5968}&		-&	mount failure\\ 
070103 &	50246&	+0.0&	2.8&	$>$19.6\,\cite{gcn5993}&	-&	mount failure\\
070129 &	50253&	+0.0&	0.4&	$>$19.2$^\dag$\cite{gcn6066}&	-&	mount failure\\
070219 &	-&	+0.0&	4.8&	$>$20.0\,\cite{gcn6111}&	-&	mount failure\\
070220 &	50258&	+0.0&	1.2&	$>$19.6\,\cite{gcn6114}&	-&	mount failure\\
070223 &	50259&	+0.0&	4.6&	$>$21.4$^\dag$\cite{gcn6134}&	-&	mount failure\\
070224 &	50260&	+1.4&	8.0&	$>$20.1$^\dag$\cite{gcn6152}&	-&	mount failure\\
070406 &	50277&	+0.0&	4.0&	-&				-&	mount failure\\
070411 &	-&	+0.0&	3.2&	$\sim$18.3\cite{gcn6288}&	-&	mount failure\\
070412 &	-&	+0.0&	3.2&	$>$20.7\cite{gcn6284}&		-&	mount failure\\
070429A&	50286&	+1.6&	1.0&	$>$18.0\cite{gcn6356}&		-&	mount failure\\
070531 &	-&	+0.0&	1.4&	$>$19.9\cite{gcn6481}&		-&	mount failure\\
070610 &	-&	+0.4&	6.2&	$\sim$19\cite{gcn6492}&		-&	mount failure\\
% \hline
070704 &	-&	+1.4&	6.2&	$>$21.2\,\cite{gcn6617}&	-&      manually disabled\\ 
070714A&	50010&	+0.0&	1.0&	-$^\dag$&			52\,s&	bad weather\\
071025&		50033&	+0.0&	0.4&	17.3\cite{gcn6992}&	-&		bad weather\\ 
071101&		50038&	+1.0&	10.6&	$>$19.7\cite{gcn7037}&	56\,s&		limit\cite{gcn7032} \\
071109&		50040&	+0.0&	1.0&	$>$15.5\cite{gcn7052}&	59\,s&		thick cirrus, limit \cite{gcn7047} \\
071112C&	50044&	+0.4&	11.0&	17.5\cite{gcn7078}&	64\,s&		bad weather\\
080320&		50076&	+0.0&	0.6&	$>$20\cite{gcn7492}&	-&		bad weather \\ 
% \hline
080330&		50079&	+0.0&	1.2&	$16.8$\cite{gcn7543}&	400\,s&		manual, detection \\
080413A&	50082&	+0.0&	1.6&	$15.0$&			60.7\,s&	detection \cite{gcn7603}\\
080430&		50090&	+0.8&	7.4&	17.5&%\cite{gcn7660}&
	34\,s&		detection \cite{gcn7648} \cite{adup080430} \\
080517&		50098&	+0.0&	2.4&	$>$18.5\cite{gcn7742}&	3\,h 22\,m&	weather delay \\
080603B&	50113&	+1.6&	6.4&	16.5\cite{gcn7808}&	1\,h 15\,m&	detection \\
080605&		50115&	+0.0&	3.8&	17.9\cite{gcn7833}&	43\,s&		detection \cite{gcn7837} \\
% \hline
080727C&	-&	+0.0&	4.8&	$>$19.9\cite{gcn8043}&	-&		dome failure\\
080903 &	-&	+0.0&	4.4&	19.2\cite{gcn8179}&	-&		dome failure\\ 
% \hline
081001&		50159&	+0.0&	1.4&	-& -& bad weather \\ % really bad wthr, we were observing with the 40cm from Cerro Burek in Argentina with Rene Duffard
081003B&	50167&	+0.0&	3.6&	$>$17.6\cite{gcn8320}&	41\,s&		OK, limit \cite{gcn8320}\\
081126&		50181&	+0.0&	4.0&	$>$18.0\cite{gcn8555}&	-&		bad weather \\ 
081128A&	50184&	+1.2&	9.8&	20.9\cite{gcn8572}&	-&		bad weather \\
081210&		50188&	+0.0&	8.0&	19.9\cite{gcn8654}&	45\,m& 		weather delay, out of focus\\
%081221&
081228&		50195&	+0.0&	2.0&	19.8\cite{gcn8752}&	-&		bad weather\\
081230&		50197&	+0.0&	3.2&	19.1\cite{gcn8754}&	-&		bad weather\\
\hline
\hline
\\
\caption{The Great Table of BOOTES-1B GRBs. "Target" is the RTS2 target number
at BOOTES-1B. $t_1$ is the time delay between the GRB trigger and the possible
start of observation. $t_\mathrm{obs}$ is the amount of time for which the GRB
can be followed until it sets for the first time. $m_{300}$ is the known
brightness of the GRB optical transient 300\,s after the event. $dT$ is the delay
of BOOTES-1B followup. 
\\{\scriptsize $\dag$ denotes discovered optical counterparts where there are not
enough data to estimate the brightness 300\,s after the GRB.} }
\label{bigtable}
\end{longtable}
}
%\end{longtable}
\end{center}

\paragraph{\it Acknowledgment.}

{\small
%We are indebted to the BOOTES-1
%Spanish-Czech team for the valuable cooperation during
%the last 10 years and to INTA, CSIC and AV\,\v{C}R for
%their support. 
The Spanish side was supported by
Spanish Ministry of Science and Technology by projects
AYA 2004-01515, AYA 2007-63677, AYA2007-67627-C03-03, AYA2008-03467/ESP and AYA2009-14000-C03-01. The Czech
participation is supported by grants of the Grant
Agency of the Czech Republic 205/08/1207, 102/09/0997
and PECS 98023.  MJ was supported by the Ministry of
Education and Science (MEC) grant AP2003-1407.}

\end{document}